\pdfoutput=1
\documentclass[a4paper, 11 pt, conference]{ieeeconf}  

\IEEEoverridecommandlockouts                              
\usepackage{amssymb}  
\usepackage[english]{babel}
\usepackage{graphicx}
\usepackage{tabularx,booktabs}
\usepackage{graphicx}
\usepackage{lipsum, babel}
\usepackage{subfigure}
\usepackage{biblatex}

\usepackage{caption} 
\renewenvironment{keywords}{
{\small	\textbf{\textit{Keywords---} Concert, Music Business, Machine Learning}
}}

\title{\LARGE \bf
A Concert-planning Tool for Independent Musicians by Machine Learning Models*
}

\author{Xiaohan Yang and Qingyin Ge  \\
 \small Department of Computer Science and Engineering \\
 \small New York University Shanghai \\
\thanks{*Thanks to Prof. Enric Junqu de Fortuny, Horace Fung, Bing and Ruowen who supported us along the way.}
\thanks{}%
\thanks{}%
}

\begin{document}

\maketitle
\pagestyle{plain}

\begin{abstract}

Our project aims at helping independent musicians to plan their concerts based on the economies of agglomeration in music industry. Initially, we planned to design an advisory tool for both concert pricing and location selection. Nonetheless, after implementing SGD linear regression and support vector regression models, we realized that concert price does not vary significantly according to different music types, concert time, concert location and ticket venues. Therefore, to offer more useful suggestions, we focus on the location choice problem by turning it to a classification task. The overall performance of our classification model is pretty good. After tuning hyper parameters, we discovered random forest gives the best performance, improving the classification result by 316\%.
This result reveals that we could help independent musicians better locate their concerts to where similar musicians would go, namely a place with higher network effects.

\end{abstract}


\section{INTRODUCTION}
Nowadays, as music streaming services develop, more and more independent artists get the opportunity to showcase their talents \cite{1}. Across genres, from hip-hop to classical, musicians are attracting large amount of fans who are willing to spend money on them, especially for concerts \cite{2}. With an increasing demand for concerts and ever-changing willingness-to-pay of the customers, independent musicians today face two challeges when planning for their concerts: how to set the ticket price and where to hold the concert.

As average ticket price has increased drastically for concerts in the US, many researches are done to explore the underlying patterns. Summarized by Appelman, the main determinants relating to the concerts are the performance quality, the artist's popularity, the venue, the stakeholders in the value chain and the music style \cite{3}. In addition, Krueger, an economist from Princeton University, suggestes that concert price also respond to economic forces since it is linked to labor market\cite{4}. Thus, we are going to explore the relationship between concert price and concert-specific factors and relative labor-market-specific factors in our regression model.

In terms of the location selection problem, it is crucial for musicians because there is a city can make music and foster stars. To explore the factors that contribute to this decision, we should not ignore economic factors since literature suggests that local economies has great impact on cultural industries, including the music industry \cite{5}. Furthermore, since influence people's taste can be influenced by cultural in different places\cite{6}, music style should also be considered in location selection process. Last but not least, as concert is not a necessity, if the concert price is not affordable for people in one place, that place is probably not a good choice. Thus, the effect of concert price should also be included in our model.

Therefore, our project will experiment on two tasks—— deciding price based on concert-specific and local-economic-specific factors, also identifying location for a given series of musician-related and economy-related features. 

\section{The Dataset and Features}
Our datasets come from SeatGeek, Last.fm and City-Data, which are all developer-friendly open sources. As a leading mobile-focused ticket platform, SeatGeek has information about over 100k upcoming entertaining events globally. Thus, we decide to use it to gather concert-related information, namely concert popularity, concert time, locations, musician music types, concert venues and price. In addition, Last.fm is one of the largest online music platform with over 21 million active users from over 200 countries, which makes it a reliable source to derive musicians popularity from. In terms of City-Data, it collects and analyzes data from a variety of government and private sources to serve over 14 million users every month. From  City-Data, we gather income per capita and population density information for every city in 2017. Based on this information, we cluster cities into 5 groups.
\par
Overall we collect information about 9,594 concerts in the US from Oct. 2018 to Dec. 2019. To train our models, we split the dataset by 0.2 factor, which means 7,675 observations in our training dataset, and 1,919 observations in our test dataset. 
\par
\begin{figure}[h]
\caption{Average price distribution}
\label{nn}
\includegraphics[width=\linewidth]{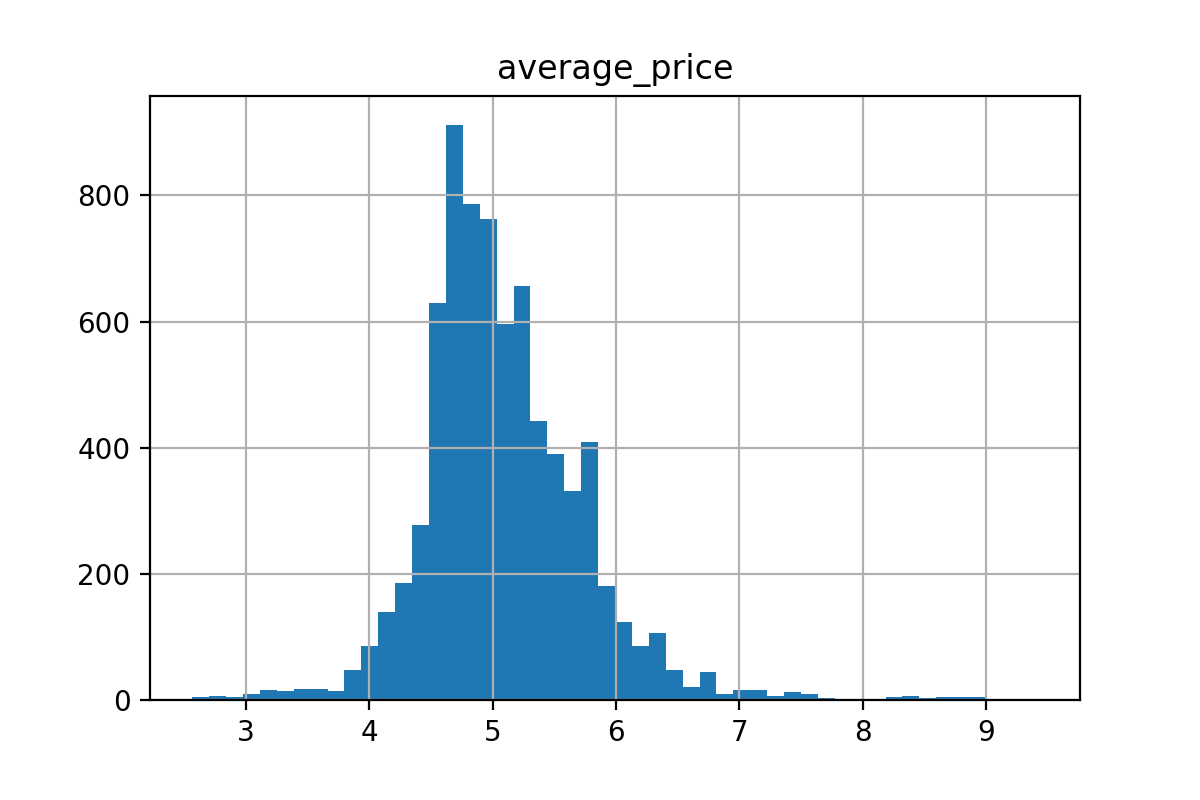}
\end{figure}

\par
When it comes to data preprocessing, we go through four steps. Firstly, we handle missing values by filling the most frequent value. In addition, we create dummies for discrete variables, namely genres and concert time. Limited by the unbalanced time window of the data collecting process, we decide not to use concert month, since it is biased. In this way,  we have 39 covariates with concert price as respondent variable for our regression problem, and 34 covariates with concert location as respondent variable for our classification problem. Moreover, as most continuing variables are either exponentially distributed or containing outliers, we use logarithm function to manipulate them. Last but not least, since our dataset contains features highly varying in magnitudes, units and range, we also use MinMax scaler to bring all the features to the same level of magnitude. 

\section{Explanation of the Method Used}

\subsection{Regression Problem}

After finishing all the preliminary data processing, we starte our training on the regression problem by using three different models. Explicitly speaking, we try Stochastic Gradient Descent Linear Regression, and Support Vector Regression to explore the pricing pattern of music concerts in the US. 

\subsubsection{SGD Linear Regression}
First and foremost, we set up a regularized linear regression with the objective to minimize the root mean squared percentage error (RMSPE), which is described as formula~\ref{eq:1} below. The reason for using RMSPE is that it describes the change in ratio: larger percentage errors, more painful. For instance, underpricing a \$30-concert by \$10 are worse than underpricing a \$100-concert by \$10. 

\begin{equation} \label{eq:1}
minimize \quad  \sqrt{\frac{1}{M}\sum_{i=1}^M (\frac{(y^{(i)}-f(x)^{(i)})}{y^{(i)}})^2} + \lambda(w)
\end{equation}

\par
As for hyperparameter tuning, we use Random Search instead of Grid Search because it is more efficient in terms of runtime. We test both L1 and L2 regularization and the result demonstrates that L1 is better. The reason behind is probably that there is collinearity among the features. Furthermore, we also tune the generalization term by searching a list of $\alpha$. Last but not least, to reduce miss\-specification errors, we also add a list of polynomial degrees in the searching model. 
\par
After Randomized Search, we determine that we should choose l2 penalty, polynomial with zero degree, regularization term $\alpha = 0.1$ as our model.

\subsubsection{Support Vector Regression}

 Support Vector Machines are supervised learning models, which can be used to solve both regression and classification problems. For this problem, we are using the formula below as a non-linear model to solve the concert pricing problem. Here, $\xi_i$ are the flexible slack variables, and C is the regularization parameter.
 $$minimize\quad \frac{1}{2}\left\Vert{w}\right\Vert^2 + \frac{1}{C}\sum_{1 = 1}^M(\xi_i + \xi_i^\prime)$$
 $$s.t.\quad y^{(i)} - (\langle w \,,\phi(x^{(i)})\rangle + w_0) \leq \varepsilon + \xi_i$$
 $$s.t.\quad y^{(i)} + (\langle w \,,\phi(x^{(i)})\rangle + w_0) \leq \varepsilon + \xi_i^\prime$$
 $$\xi_i, \xi_i^\prime \geq 0$$
 
 For the sake of consistency, RMSPE is still used as the loss function. Since radial basis function (RBF)  is recommended by some literature, we use RBF to conduct kernel trick.
 $$\vec{w}\cdot \phi(\vec{x}) = \sum_{i}\alpha_{i}y_{i}\exp(-\gamma\left\Vert{\vec{x}_{i} - \vec{x}_{j}}\right\Vert^2)$$
 The main hyper-parameters for SVM Regression are the contribution of the i-th training data point to the final solution w, RBF coefficient $\gamma$, and the margin of error tolerance $\varepsilon$. In order to avoid over-fitting, we choose small C to lower the model complexity and small $\gamma$ to restrict the region of influence of a single SV. Furthermore, we set a list of large $\varepsilon$ to include more observations in the tolerance range.

After applying Grid Search, the model is specified as follows:
$C = 0.5$,
$\gamma = 0.01$,
$\varepsilon = 2$
\subsection{Classification Problem}
After solving the regression problem, we try to identify locations for different concerts. As stated before, we cluster five classes for cities by k-means, based on their income per capita and population density. Therefore, our problem is actually building a model to identify which class of cities a concert should take place. Nevertheless, since the weight of different class are not the same, oversampling is used to balance the training dataset. Probing into this problem, we test four models, which are SGD Logistic regression, Support Vector Classifier, Neural Networks and Random Forest.

\subsubsection{Logistic Regression}
First, we start with logistic regression since it is the most basic and simple method in classification problems. Our formula is as follows, where C is the inverse of regularization strength.

$$f(x) = \frac{1}{1+e^{-(w_0 + w^T X)}} + \frac{1}{C}\left\Vert{w}\right\Vert$$

As 26 of our independent variables are dummies, there is a high potential that we can compress the information into a lower-dimension matrix. Thus, we add Principal Component Analysis (PCA) in the pipeline to test if by reducing the dimension of vector space, the benefit of lowering variance could overweight the loss of information. Furthermore, with five classes in the model, we test both one-versus-rest loss fit and multinomial loss fit for each label. Moreover, to reduce over-fitting errors, a regularization term is added. 
After tuning all the hyperparameters by using Randomized Search, we decide to set $C$ to be 0.1 and to use L1 penalty. This following function does not include the PCA process because the searching result indicates that the benefit of reducing dimension is not satisfying.

\subsubsection{Support Vector Machines}
Other than logistic regression, we implement a SVM Classifier with RBF kernel \cite{7} to solve this problem. Similar to Support Vector Regression, regularization parameter C, RBF coefficient $\gamma$ are crucial hyperparameters for us to tune. To avoid over-fitting problems, we constrain C and $\gamma$ in a small range to limit the model complexity. After tuning hyper parameters by applying Randomized Search, we decide to set $C$ to be 10 and $\gamma $ to be 0.01. The referred function is as follows.
 $$ f(x) = \sum_{i = 1}^{M}y^{(i)}\alpha^{(i)}*\langle\phi(x^{(i)})\,,\phi(x^{(i)})\rangle_\mathcal{H} + w_0 $$
 From here we can just optimize this function without intensive calculation \cite{c8}.
 $$maximize \quad\sum_{i = 1}^{M}\alpha_i - \frac{1}{2}\sum_{i,j = 1}^{M}y^{(i)}y^{(j)}\alpha^{(i)}\alpha^{(j)}\kappa(x^{(i)},x^{(j)})$$
 $$subject\: to \quad \sum_{i = 1}^{N} y_i\alpha_i = 0$$
 $$where \quad 0 \leq \alpha_i \leq \frac{1}{\lambda}$$

\subsubsection{Neural Networks}
Since we want a better result, we try to use neural network to improve the accuracy. Since by neural network method, we can tune more parameters, and also by using more layers and layer with more nodes we can filter features thereby generate a better result. However, if not careful enough, there is a big chance for us to over-fit our data, and we actually encounter with this problem. We set the baseline to be a model without any regularization, setting epoch = 1000. This baseline that contains four layers is fully over-fitting: three input layers with 64, 16, and 16 neurons and one output layer with 5 neurons. Except for the last layer, we choose ReLU function since it generally can make the whole neural network system more stable. Since we are solving multiclass classification problem, we use Softmax as our last activation function. To avoid over-fitting and get the best model, we use Keras together with grid search, add more restrictions and dropouts, also decrease the number of layers and neurons to restrain the model.

\subsubsection{Random Forest}
Random Forest is also a good method to conduct a trial, since it is considered as a very handy and easy algorithm, and it’s default hyperparameters often produce a good prediction result. In addition, since it is a bagging method, specifically it contains several small decision trees, we think it should be robust and accurate enough. Because of these convenience, we just build a simple random forest classifier using sklearn, and tune mainly three parameters: one is number of trees, one is the minimum number of samples required to be at a leaf node and the other one is the maximum depth of the tree. Here we use random search, randomly picking number from proper range and see which one gives the best result.

\section{Results}
 To evaluate the foregoing models, we set various benchmarks. Explicitly speaking, the RMSPE of the constant model of mean value is used as the measuring standard for the regression problem. Moreover, for the classification problem, the accuracy of a random guess is used as the lower bound and the accuracy of extremely over-fitting model is used as the upper bound. The testing results reveals that the constant model performs the best for the regression problem and the random forest performs the best for the classification problem. 
 \par
 \begin{table}[h]
\caption{Regression Results}
\label{rr}
\label{result1}
\begin{center}
\resizebox{\linewidth}{!}{%
\begin{tabular}{@{}llcc@{}}
\toprule
            & Benchmark & \multicolumn{1}{l}{SGD Linear Regression} & \multicolumn{1}{l}{Support Vector Regression} \\ \midrule
Traning RMSPE & 0.131     & 0.131                                     & 0.156                                         \\
Testing RMSPE & 0.133     & 0.132                                     & 0.161                                         \\ \bottomrule
\end{tabular}%
}
\end{center}
\end{table}

Probing into the regression result, the linear model and SVM regression do not improve the benchmark value significantly. As shown in Table \ref{rr}, baseline indicates that RMSPE is 0.133, SGD Linear Regression has RMSPE 0.132, and SVM yields 0.161. We can see that constant function does the best job and SVM does worst. There are probably two reasons can be identified. First and foremost, since our benchmark result gives us a constant function, so it is possible that only by using constant function we can fit our data with enough complexity and lower variance. When applying SGD linear regression model we are getting similar result: it learns polynomial degree 0 as the best fit. Therefore, by increasing model complexity, in other words, using SVM regression, the result is getting worse. Probably because we are over-fitting and increase the variance, but bias doesn’t change significantly. Second of all, there might be no strong relationship between our covariates and respondent variable, or we do not tune our parameters perfectly, which leads to worse result. We could discuss further if time is permitted.

\begin{table}[h]
\caption{Classification Results}
\label{cr}
 
\begin{center}
\resizebox{\linewidth}{!}{%
\begin{tabular}{lcccc}
\toprule
                  & \multicolumn{1}{l}{Logistic Regression} & \multicolumn{1}{l}{SVM} & \multicolumn{1}{l}{Neural Networks} & \multicolumn{1}{l}{Random Forest} \\ \midrule
Benchmark (Low)   & 20\%                                    & 20\%                    & 20\%                                & 20\%                              \\
Benchmark (High)  & 48.8\%                                  & 49.3\%                  & 73.4\%                              & 100\%                                \\
Training Accuracy & 45.2\%                                  & 45.8\%                  & 57.0\%                                  & 52.2\%                                \\
Testing Accuracy  & 48.3\%                                  & 49.7\%                     & 52.1\%                                  & 63.1\%                               \\ \bottomrule
\end{tabular}%
}
\end{center}
\end{table}

Despite of the constant pattern in concert pricing, our classification results demonstrates that concert location is significantly influenced by cities’ economic classes. Based on income per capita and population density, cities are clustered into five classes. 

With five classes in total, we set our accuracy lower bound to be 20\%, namely the random guess accuracy. As shown in Table \ref{cr}, our best models improve the testing accuracy by 310\%, which verifies the existence of location-selection patterns for concerts in the U.S. Furthermore, comparing with the upper bound of each model, our testing result reaches the 70\% capacity on average. Among all the models, Neural Networks and Random Forest perform better.

\begin{figure}[h]
\caption{Neural Network Diagram}
\label{nn}
\includegraphics[width=\linewidth]{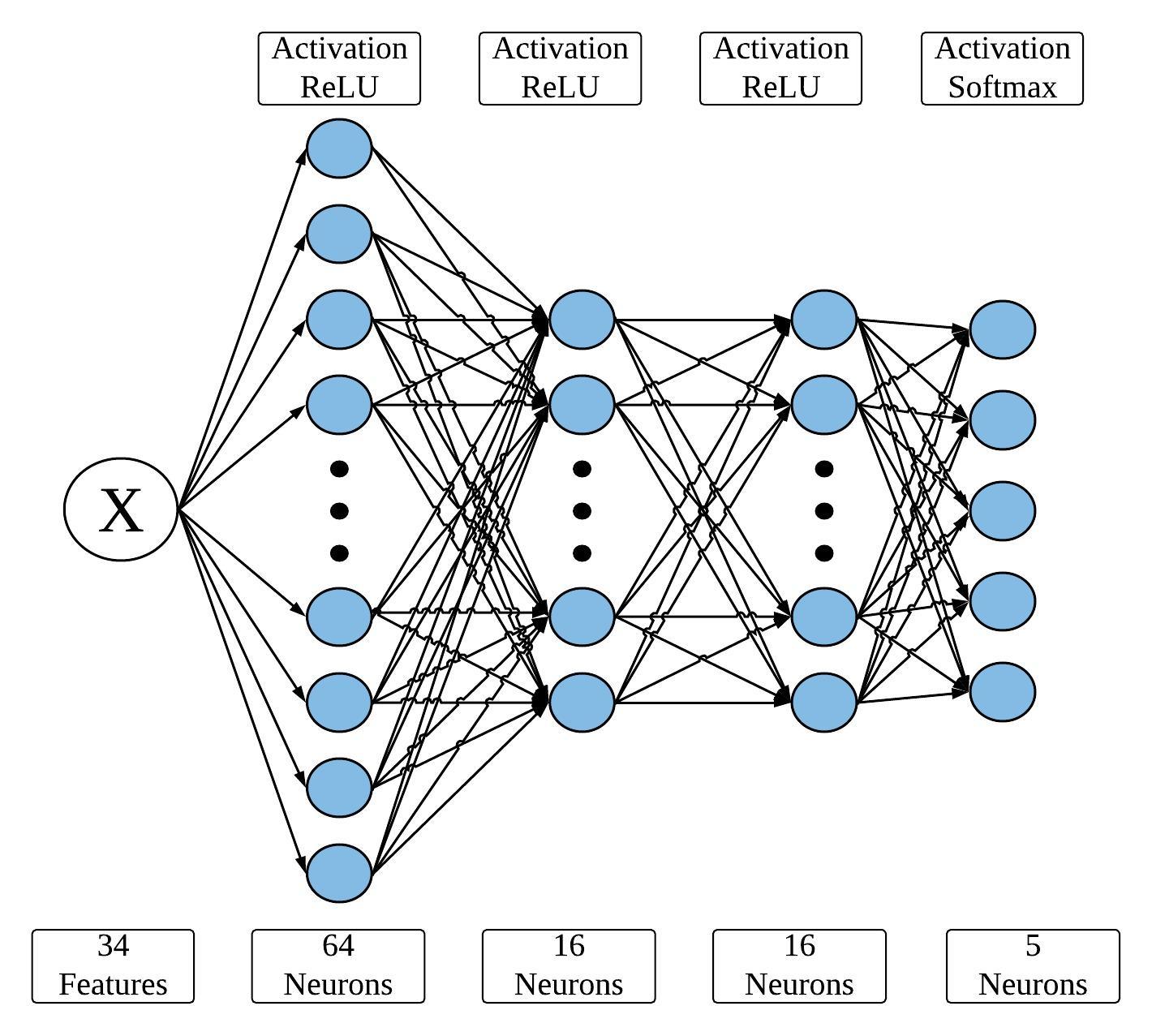}
\end{figure}
\par
 Our baseline here, without regularizing anything, we obtain 73.4\% accuracy, and the best model we test for neural networks is presented in Figure \ref{nn}. We can see that here we have totally four layers with 64, 16, 16, and 5 neurons on each layer, respectively. 
\begin{figure}[h]
\caption{Accuracy Training Process}
\label{at}
\includegraphics[width=\linewidth]{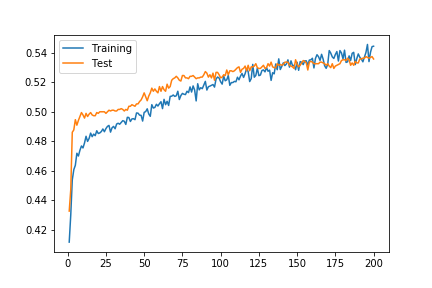}
\end{figure}
\par
By setting the activation as ReLU for the first three layers and softmax for the last layer, the training and testing accuracy changes as in Figure \ref{at}. As epochs increases, the accuracy growth for testing set slows down. Finally, we obtain the result with 52.3\% accuracy within 200 epochs. 

\begin{figure}[h]
\caption{Plain and Normalized Confusion Matrix}
\label{pncm}
\includegraphics[width=\linewidth]{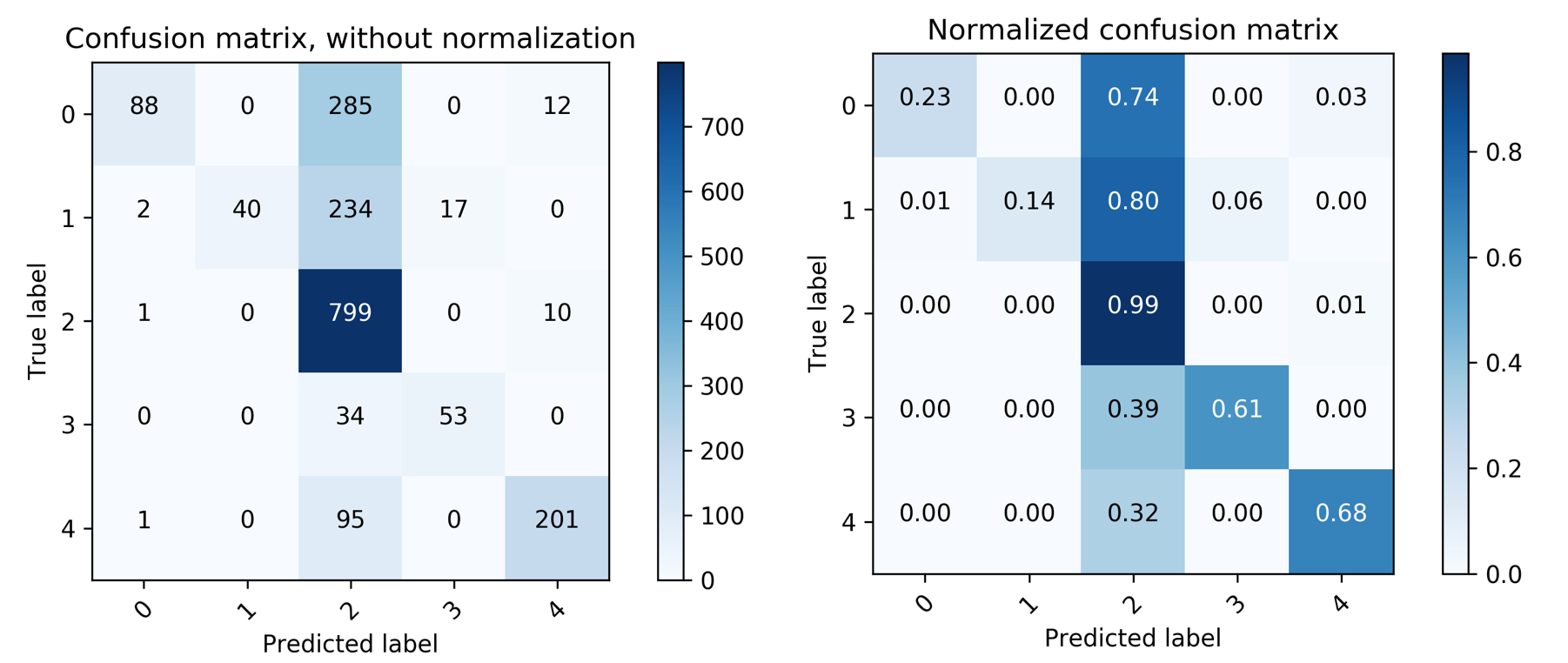}
\end{figure}

In addition, Random Forest also provides us a good result in classification. For this model, we set the higher bound to be 100\% with the minimum number of samples equals 1. Namely, this is the extremely over-fitting accuracy. By applying Randomized Search, the best parameters for our model are 105, 47 and 10 for number of trees, max depth, and minimum number of samples respectively. The predicting accuracy of this model is 63.1\% and the confusion matrices are presented in Figure \ref{pncm}.
\par
Last but not least, logistic regression and SVM classification does not perform as well as the others because of the model limitation. The testing results of these models are around 45.5\%, which is 92.9\% of the higher bound, namely the result of over-fitting training data.

\section{Conclusion and Future Work}
In conclusion, our work contribute to the concert planning process for independent musicians from pricing and location selection perspectives.
To begin with, we discover that concert prices always fluctuate in a certain range regardless of different musician-specific and local-economy-specific features. As suggested by other researchers, concert is an "experience good", which people need to attend to know the utility \cite{8}. Since the price is also one of the experiment factor, musicians usually prefer not to charge an unaffordable price to differentiate themselves \cite{9}. Therefore, it is understandable that a constant function with $y = 160.774$ generates the lowest average absolute error. 
\par
When it comes to concert location classification, random forest and neural network yield best performance. It is as expected that the bagging method and method involve deep learning have higher accuracy. We improve the random result (baseline = 20\%) by almost triples, 63\% \& 53.5\% respectively, which is already a huge jump. We are now be able to help musicians plan their concert location to proper city class with 60\% accuracy, so they could eventually meet customers expectation in some ways. 
\par 
Looking forward, we can improve our model by adding more observations and more features. Firstly, limited by our data source, we only know the concerts that are going to happen soon, but we have little information about historical concerts. After gathering a more balanced data set, we could explore the pattern of concert time and improve our current models with lower bias. Furthermore, we could also add more musician-charisma-related and the concert-profit-related features, since the profitability pattern of concerts is not discussed in this paper.

\clearpage

\addtolength{\textheight}{-12cm}   




\onecolumn
\pagestyle{empty}
\section{APPENDIX}
\begin{table}[h]
\caption{Variable Specification}
\resizebox{\textwidth}{!}{%
\begin{tabular}{@{}ll@{}}
\toprule
Feature                                              & Data Description                                                                                                         \\ \midrule
average price                                        & The respondent variable in price prediction process.                                                                     \\
latitude                                             & Latitude of concert city                                                                                                 \\
longitude                                            & Longitude of concert city                                                                                                \\
concert\_popularity                                  & A numerical representation of popularity based on ticket sales from SeatGeek                                             \\
playcount                                            & How many times this musician's song have been displayed                                                                  \\
Population\_Estimate\_2017                           & 2017 American city population. A feature to determine city class.                                                        \\
market\_heat                                         & How many concerts will be held in this city.                                                                             \\

Estimated\_per\_capita\_income                       & Income per capita. A feature to determine price.                                                                         \\
Population\_density                                  & Population density of a city. A feature to determine city class.                                                         \\

Class                                                & After depending on all the city features, the city class be clustered. The respondent variable in classification problem \\
alternative                                          & Dummy. Music Type.                                                                                                       \\
blues                                                & Dummy. Music Type.                                                                                                       \\
classic-rock                                         & Dummy. Music Type.                                                                                                       \\
classical                                            & Dummy. Music Type.                                                                                                       \\
country                                              & Dummy. Music Type.                                                                                                       \\
electronic                                           & Dummy. Music Type.                                                                                                       \\
folk                                                 & Dummy. Music Type.                                                                                                       \\
hip-hop                                              & Dummy. Music Type.                                                                                                       \\
hard-rock                                            & Dummy. Music Type.                                                                                                       \\
indie                                                & Dummy. Music Type.                                                                                                       \\
jazz                                                 & Dummy. Music Type.                                                                                                       \\
latin                                                & Dummy. Music Type.                                                                                                       \\
punk                                                 & Dummy. Music Type.                                                                                                       \\
pop                                                  & Dummy. Music Type.                                                                                                       \\
rap                                                  & Dummy. Music Type.                                                                                                       \\
reggae                                               & Dummy. Music Type.                                                                                                       \\
rnb                                                  & Dummy. Music Type.                                                                                                       \\
rock                                                 & Dummy. Music Type.                                                                                                       \\
soul                                                 & Dummy. Music Type.                                                                                                       \\
techno                                               & Dummy. Music Type.                                                                                                       \\
genres\_num                                          & How many music types a musician possess.                                                                                 \\
venue\_concert\_count                                & How many concerts a venue support.                                                                                       \\
venue\_type                                          & Large, medium or small venue.                                                                                            \\
Sun                                                  & Dummy. Concert performance day.                                                                                          \\
Mon                                                  & Dummy. Concert performance day.                                                                                          \\
Tue                                                  & Dummy. Concert performance day.                                                                                          \\
Wed                                                  & Dummy. Concert performance day.                                                                                          \\
Thu                                                  & Dummy. Concert performance day.                                                                                          \\
Fri                                                  & Dummy. Concert performance day.                                                                                          \\
Sat                                                  & Concert performance day.                                                                                                 \\ \bottomrule
\end{tabular}%
}
\end{table}

\clearpage

\begin{table}[h]
\caption{Data Description Without Dummies}
\resizebox{\textwidth}{!}{%
\begin{tabular}{lrrrrrrrr}

\toprule
{} &   count &        mean &          std &        min &         25\% &         50\% &         75\% &          max \\
\midrule
average\_price                                 &  7472.0 &       5.090 &        0.655 &      2.565 &       4.673 &       4.997 &       5.438 &        9.418 \\
latitude                                      &  7472.0 &      38.176 &        5.295 &     10.000 &      34.000 &      39.000 &      42.000 &       61.000 \\
longitude                                     &  7472.0 &     -94.828 &       17.695 &   -158.000 &    -115.000 &     -89.000 &     -80.000 &        9.000 \\
concert\_popularity                            &  7472.0 &       0.511 &        0.090 &      0.000 &       0.440 &       0.490 &       0.550 &        0.880 \\
playcount                                     &  7472.0 &      15.096 &        1.666 &      9.536 &      13.991 &      15.079 &      16.324 &       19.428 \\
Population\_Estimate\_2017                      &  7472.0 &  974887.883 &  1607633.700 &  47929.000 &  187347.000 &  486290.000 &  879170.000 &  8622698.000 \\
market\_heat                                   &  7472.0 &     207.377 &      172.505 &      1.000 &      57.000 &     155.000 &     346.000 &      568.000 \\
Estimated\_per\_capita\_income                   &  7472.0 &   27178.475 &     7233.040 &  14221.000 &   22180.750 &   26159.000 &   30136.600 &    51686.000 \\
Population\_density                            &  7472.0 &    5552.186 &     5309.790 &    668.000 &    2594.000 &    3574.000 &    6237.000 &    27714.000 \\
Class                                         &  7472.0 &       1.804 &        1.291 &      0.000 &       1.000 &       2.000 &       2.000 &        4.000 \\
genres\_num                                    &  7472.0 &       0.930 &        0.425 &      0.000 &       0.693 &       1.099 &       1.386 &        1.792 \\
venue\_concert\_count                           &  7472.0 &       2.498 &        0.902 &      0.000 &       1.946 &       2.639 &       3.135 &        3.970 \\
venue\_type                                    &  7472.0 &       1.943 &        0.861 &      1.000 &       1.000 &       2.000 &       3.000 &        3.000 \\

\bottomrule

\end{tabular}%
}

\end{table}


\clearpage


\begin{thebibliography}{99}

\bibitem{1} Khalili-Tari, D. (2017, December 15). How independent artists have changed the music industry. Retrieved from https://www.independent.co.uk/arts-entertainment/music/features/independent-artists-music-industry-stormzy-aj-tracey-stefflon-don-hardy-caprio-major-label-streaming-a8110936.html

\bibitem{2}
Gervais, A. (2015, December 26). Why Musicians Aren't Paid More Fairly. Retrieved from https://aarongervais.com/blog/musicians-arent-paid/

\bibitem{3}
Decrop, A., and Derbaix, M. (2014). Artist-Related Determinants of Music Concert Prices. Psychology \& Marketing, 31(8), 660-669. doi:10.1002/mar.20726

\bibitem{4}
Alan B. Krueger, "The Economics of Real Superstars: The Market for Rock Concerts in the Material World," Journal of Labor Economics 23, no. 1 (January 2005): 1-30.
https://doi.org/10.1086/425431

\bibitem{5} Kloosterman, R. (2005). Come Together: An Introduction to Music and the City. Built Environment (1978-), 31(3), 180-191. Retrieved from http://www.jstor.org/stable/23289438
\bibitem{6} Harrington, H. (2005). Black Stars or Black Holes? Cities as Sites for Verticality in Popular Black Music Production. Built Environment (1978-), 31(3), 208-225. Retrieved from http://www.jstor.org/stable/23289440


\bibitem{7} Amari, S., Wu, S. (1999). Improving support vector machine classifiers by modifying kernel functions. NEURAL NETWORKS -OXFORD-, (6), 783. Retrieved from http://search.ebscohost.com/login.aspx?direct=true

\bibitem{8} Madzarov, Gjorgji, et al. A Multi-Class SVM Classifier Utilizing Binary Decision Tree. 27 July 2008.

\bibitem{9}Courty, P. (2000). An economic guide to ticket pricing in the entertainment industry. Recherches Économiques De Louvain / Louvain Economic Review, 66(2), 167-192. Retrieved from http://www.jstor.org.proxy.library.nyu.edu/stable/40724285







\end{thebibliography}
\end{document}